\documentclass[11pt]{article}
\usepackage{amsmath}  
\begin{document}
\renewcommand{\theequation}{\thesection.\arabic{equation}}
\def\prg#1{\medskip{\bf #1}}     \def\Ra{\Rightarrow}
\def\lra{\leftrightarrow}        \def\Lra{{\Leftrightarrow}}
\def\nind{\noindent}             \def\pd{\partial}
\def\dis{\displaystyle}          \def\dfrac{\dis\frac}
\def\grl{{GR$_{\Lambda}$}}       \def\vsm{\vspace{-10pt}}
\def\Leff{\hbox{$\mit\L_{\hspace{0.6pt}\rm eff}\,$}}
\def\egt{{\scriptstyle\rm EGT}}  \def\cs{{\sc cs}}

\def\cL{{\cal L}}     \def\cM{{\cal M }}    \def\cK{{\cal K}}
\def\cO{{\cal O}}     \def\cE{{\cal E}}     \def\cH{{\cal H}}
\def\cC{{\cal C}}     \def\tR{{\tilde R}}   \def\tG{{\tilde G}}
\def\tI{{\tilde I}}   \def\bD{{\bar D}}     \def\bH{{\bar H}}
\def\bI{{\bar I}}     \def\hO{{\hat O}}     \def\hcO{\hat{\cO}}
\def\hcH{\hat{\cH}}   \def\bcH{{\bar\cH}}   \def\bcK{{\bar\cK}}
\def\Ir{I_{\rm reg}}  \def\tIr{{\tilde\Ir}}

\def\G{\Gamma}        \def\S{\Sigma}
\def\D{{\Delta}}      \def\L{{\mit\Lambda}}
\def\a{\alpha}        \def\b{\beta}         \def\g{\gamma}
\def\d{\delta}        \def\m{\mu}           \def\n{\nu}
\def\th{\theta}       \def\k{\kappa}        \def\l{\lambda}
\def\vphi{\varphi}    \def\ve{\varepsilon}  \def\p{\pi}
\def\r{\rho}          \def\Om{\Omega}       \def\om{\omega}
\def\s{\sigma}        \def\t{\tau}          \def\tom{{\tilde\omega}}
\def\nn{\nonumber}
\def\be{\begin{equation}}             \def\ee{\end{equation}}
\def\ba#1{\begin{array}{#1}}          \def\ea{\end{array}}
\def\bea{\begin{eqnarray} }           \def\eea{\end{eqnarray} }
\def\beann{\begin{eqnarray*} }        \def\eeann{\end{eqnarray*} }
\def\beal{\begin{eqalign}}            \def\eeal{\end{eqalign}}
\def\lab#1{\label{eq:#1}}             \def\eq#1{(\ref{eq:#1})}
\def\bsubeq{\begin{subequations}}     \def\esubeq{\end{subequations}}
\def\bitem{\begin{itemize}}           \def\eitem{\end{itemize}}

\title{Covariant description of the black hole entropy in 3D gravity}

\author{M.\ Blagojevi\'c and B. Cvetkovi\'c\footnote{
        Electronic addresses: mb@phy.bg.ac.yu, cbranislav@phy.bg.ac.yu}\\
\normalsize Institute of Physics, P.O.Box 57, 11001 Belgrade, Serbia}
\date{}
\maketitle

\begin{abstract}
We study the entropy of the black hole with torsion using the covariant
form of the partition function. The regularization of infinities
appearing in the semiclassical calculation is shown to be consistent
with the grand canonical boundary conditions. The correct value for the
black hole entropy is obtained provided the black hole manifold has two
boundaries, one at infinity and one at the horizon. However, one can
construct special coordinate systems, in which the entropy is
effectively associated with only one of these boundaries.
\end{abstract}


\section{Introduction} 
\setcounter{equation}{0}

Three-dimensional (3D) gravity is an attractive model for
investigating basic features of both classical and quantum gravity.
In the traditional approach based on general relati\-vi\-ty (GR),
gravitational dynamics is studied in spacetime with an underlying
{\it Riemannian\/} structure \cite{1,2,3,4,5,6,7,8,9}. On the other
hand, it has been well known for nearly five decades that there
exists a gauge-theoretic conception of gravity, based on {\it
Riemann-Cartan\/} geometry of spacetime (see, e.g. \cite{10,11,12}).
In this approach, both the {\it curvature\/} and the {\it torsion\/}
are used to characterize the gravitational dynamics. The application
of these ideas to 3D gravity started in the early 1990s, leading to a
deeper understanding of the dynamical role of torsion
\cite{13,14,15,16,17,18,19,20,21,22,23}.

In 1991, Mielke and Baekler proposed a topological model for 3D
gravity based on Riemann-Cartan geometry \cite{13}. The related field
equations have one particularly interesting solution---the black hole
with torsion \cite{15,16,17,18,19},  which generalizes the BTZ black
hole \cite{4}. It is well known that quantum nature of gravity is
reflected in {\it thermodynamic\/} properties of black holes
\cite{24,25,26,27,28}. In the standard field-theoretic approach,
these properties can be described by the {\it Euclidean\/} functional
integral \cite{4,6}:
\be
Z[\b,\Om]=\int Db^iD\om^j
          \exp\left(-\tI[b^i,\om^j,\b,\Om]\right)\, .      \lab{1.1}
\ee
Here, $b^i$ and $\om^j$ are the triad field and the connection, $\b$
and $\Om$ are the Euclidean time period and the angular velocity of
the black hole, and $\tI$ is the Mielke-Baekler action, corrected by
suitable boundary terms. The boundary conditions are chosen so that
$Z[\b,\Om]$ is the grand canonical partition function. Using the {\it
Hamiltonian\/} form of the action and assuming that the black hole
manifold has {\it two boundaries\/}, one at infinity ($B^\infty$) and
one at the horizon ($B^{r_+}$), one can calculate the entropy of the
black hole with torsion \cite{22} (for a derivation based on the
Cardy formula, see \cite{23}). The result is found to be consistent
with the first law of black hole thermodynamics.

One could expect that working with the Hamiltonian form of the action
is not of particular importance, and that transition to the {\it
covariant\/}, Lagrangian form represents only a technical step, which
cannot change the final result for the entropy. However, here we have
at least two issues that deserve a careful analysis. (a) In the
semiclassical calculation of the partition function \eq{1.1}, one needs
the value of the covariant action at the black hole configuration,
which is a divergent expression. The existence of this divergence can
be taken care of by a convenient regularization procedure, but one
should verify that this \hbox{procedure} is consistent with the adopted
boundary conditions \cite{29}. (b) Although one expects a boundary at
the horizon in the Hamiltonian formalism \cite{4,28}, it seems that its
presence in the covariant formalism of GR can be ignored \cite{29}.
This situation needs a consistent explanation.

The purpose of the present work is to clarify the role of the
boundaries $B^\infty$ and $B^{r_+}$ in the covariant description of
the black hole entropy, based on the Mielke-Baekler action. In
particular, we show that separate contributions stemming from
$B^\infty$ and $B^{r_+}$ do not have an invariant meaning---they
depend on the coordinate system used in the calculations, while the
invariant physical content can be ascribed only to the complete
boundary $B^\infty\cup B^{r_+}$.

The layout of the paper is as follows. In Sect. II, we present basic
aspects of the Euclidean 3D gravity with torsion, including the form
of the black hole solution. In Sect. III, we demonstrate the
consistency of the regularization procedure with the grand canonical
boundary conditions. Using the standard, ``rotating" coordinate
system \cite{6,29}, we calculate the covariant grand canonical
partition function \eq{1.1} and obtain the correct expression for the
black hole entropy, provided the black hole manifold has {\it two
boundaries\/}, one at infinity and one at the horizon. Both of these
boundaries give nontrivial contributions to the black hole entropy.
In Sect. IV, these considerations are extended to a more general
class of coordinate systems. While the entropy remains unchanged, as
one expects, we find that there exists a particular coordinate system
in which the contribution stemming from $B^{r_+}$ vanishes, and
consequently, $B^{r_+}$ becomes irrelevant and can be ignored.
Similar construction is then carried out for $B^\infty$. In these
coordinate systems, the complete black hole entropy is effectively
associated with a {\it single boundary\/}---either $B^\infty$ or
$B^{r_+}$. Sect. V is devoted to concluding remarks, while appendices
contain some technical details.

Our conventions are the same as in Ref. \cite{22}: the Latin indices
$(i,j,k,...)$ refer to the local orthonormal frame, the Greek indices
$(\m,\n,\r,...)$ refer to the coordinate frame, and both run over
$0,1,2$; $\eta_{ij}=(+,+,+)$ are metric components in the local
frame; totally antisymmetric tensor $\ve^{ijk}$ and the related
tensor density $\ve^{\m\n\r}$ are both normalized by $\ve^{012}=+1$.

\section{Euclidean 3D gravity with torsion} 
\setcounter{equation}{0}

Euclidean 3D gravity with torsion can be formulated as a gauge theory
of the Euclidean group $ISO(3)$ \cite{22}. In this approach, basic
dynamical variables are the triad field $b^i$ and the spin connection
$\om^i$ (1-forms), and the corresponding field strengths are the
torsion and the curvature (2-forms):
$T^i:= db^i+\ve^i{}_{jk}\om^j\wedge b^k$,
$R^i:=d\om^i+\frac{1}{2}\,\ve^i{}_{jk}\om^j\wedge\om^k$.
The geometric structure of $ISO(3)$ gauge theory corresponds to {\it
Riemann-Cartan\/} geometry \cite{10,11,12}.

\subsection{The action integral}

Mielke and Baekler proposed a {\it topological\/} model for 3D
gravity in Riemann-Cartan spacetime \cite{13}, which is a natural
generalization of Riemannian GR with a cosmological constant (\grl).
Euclidean version of the model is defined by the action $I_E$,
obtained from its Minkowskian counterpart $I_M$ by the process of
analytic continuation $I_M\to iI_E$ \cite{22}. Omitting the
subscript $E$ for simplicity, the Euclidean Mielke-Baekler action
reads
\bsubeq\lab{2.1}
\be
I=aI_1+\L I_2+\a_3I_3+\a_4I_4+I_m\, ,                      \lab{2.1a}
\ee
where
\bea
&&I_1= 2\int b^i\wedge R_i\, ,                             \nn\\
&&I_2=-\frac{1}{3}\,\int\ve_{ijk}b^i\wedge b^j\wedge b^k\,,\nn\\
&&I_3=\int\left(\om^i\wedge d\om_i
  +\frac{1}{3}\ve_{ijk}\om^i\wedge\om^j\wedge\om^k\right)\,,\nn\\
&&I_4=\int b^i\wedge T_i\, ,                               \lab{2.1b}
\eea
\esubeq
and $I_m$ is a matter contribution. The first two terms are of the
same form as in \grl, $a=1/16\pi G$ and $\L$ is the cosmological
constant, $I_3$ is the Chern-Simons action for the connection,
and $I_4$ is a torsion counterpart of $I_1$.

In the sector $\a_3\a_4-a^2\ne 0$, the vacuum field equations are
non-degenerate:
\bsubeq\lab{2.2}
\be
2T^i=p\ve^i{}_{jk}\,b^j\wedge b^k\, ,\qquad
2R^i=q\ve^i{}_{jk}\,b^j\wedge b^k\, ,                      \lab{2.2a}\\
\ee
with
\be
p:=\frac{\a_3\L+\a_4 a}{\a_3\a_4-a^2}\, ,\qquad
q:=-\frac{(\a_4)^2+a\L}{\a_3\a_4-a^2}\, .                  \lab{2.2b}
\ee
\esubeq
Note that $p$ and $q$ satisfy the identities
\be
aq+\a_4 p-\L\equiv 0\, ,\qquad ap+\a_3q+\a_4\equiv 0\, .   \lab{2.3}
\ee

Introducing the Levi-Civita connection $\tom^i$ by
$db^i+\ve^i{}_{jk}\tom^j b^k=0$, one can use the field equations to
find the Riemannian piece of the curvature $R^i(\tom)$ \cite{22}:
\be
2R^i(\tom)=\Leff\,\ve^i{}_{jk}\,b^j\wedge b^k\, ,\qquad
\Leff:= q-\frac{1}{4}p^2\, ,                               \lab{2.4}
\ee
where $\Leff$ is the effective cosmological constant. Thus, our
spacetime is maximally symmetric with isometry group $SO(3,1)$, and
it is known as the hyperbolic 3D space $H^3$.
In what follows, we restrict our attention to the Euclidean
continuation of anti-de Sitter space, which is defined by positive
$\Leff$: $\Leff=:{1}/{\ell^2}>0$.

\subsection{The black hole with torsion}

For $\Leff>0$, equation \eq{2.4} has a well-known solution for the
metric, the Euclidean BTZ black hole \cite{4,6}. In Schwarzschild
coordinates $x^\m=(t,r,\varphi)$, the metric has the form
\bea
&&ds^2=N^2dt^2+N^{-2}dr^2+r^2(d\varphi+N_\vphi dt)^2\, ,   \nn\\
&&N^2=\left(-8Gm+\frac{r^2}{\ell^2}-\frac{16G^2J^2}{r^2}\right),
  \qquad N_\vphi=-\frac{4GJ}{r^2}.                         \lab{2.5}
\eea
The zeros of $N^2$, $r_+$ and $r_-=-i\r_-$, are related to the black
hole parameters $m$ and $J$ by the relations
$r^2_+-\r^2_-=8Gm\ell^2, r_+\r_-=4GJ\ell$, both $\vphi$ and $t$
are periodic:
$$
0\le\vphi<2\pi\, ,\qquad 0\le t<\b\, ,\qquad
         \b=\frac{2\pi\ell^2 r_+}{r_+^2+\r_-^2}\, ,
$$
and the black hole manifold is topologically a solid torus
\cite{6,27,28}.

Starting with the BTZ metric \eq{2.5}, one can find the pair ($b^i$,
$\om^i$) which solves the field equations \eq{2.2}, and represents
the Euclidean black hole with torsion \cite{22}. Energy and angular
momentum of the solution are
\be
E= m+\frac{\a_3}{a}\left(\frac{pm}{2}-\frac{J}{\ell^2}\right),
\quad M= J+\frac{\a_3}{a}\left(\frac{pJ}{2}+m\right)\, .   \lab{2.6}
\ee

Instead of using the Schwarzschild coordinates, we shall go over to a
new class of coordinate systems, which is suitable for exploring the
geometric origin of the black hole entropy. Let us first introduce the
``rotating'' coordinate system, denoted by $K_0$, by
$$
t':=t/\b\, ,\qquad\vphi':=\vphi+\Om t\, ,
$$
where $\Om:=N_\vphi(r_+)=-{\r_-}/{\ell r_+}$ is related to the
angular velocity of the black hole, and $\vphi'$ it the usual
azimuthal angle  \cite{6,29}. Our new class of coordinate systems
$K_w$ is defined as a simple generalization of $K_0$:
\be
t'':=t'+w\vphi'\, ,\qquad \vphi'':=\vphi'\, ,              \lab{2.7}
\ee
where $w=w(\b,\Om)$ is a parameter. Ignoring double primes for
simplicity, the black hole solution $(b^i,\om^j)$ in $K_w$ takes the
form
\bsubeq\lab{2.8}
\bea
&&b^0=\b N(dt-wd\vphi)\, ,\qquad b^1=N^{-1}dr\, ,          \nn\\
&&b^2=r\left[d\vphi+\b(N_\vphi-\Om)(dt-wd\vphi)\right]\, , \lab{2.8a}\\
&&\om^i=\tom^i+\frac{p}{2}\,b^i\, ,                        \lab{2.8b}
\eea
where the Levi-Civita connection $\tom^i$ is:
\bea
&&\tom^0= N[d\vphi-\b\Om (dt-wd\vphi)]\, ,\qquad
  \tom^1=-N^{-1}N_\vphi dr\, ,                             \nn\\
&&\tom^2=-\b\left(\frac{r}{\ell^2}+rN_\vphi\Om\right)(dt-wd\vphi)
         + rN_\vphi d\vphi\, .                             \lab{2.8c}
\eea
\esubeq

\section{The black hole entropy in {\boldmath $K_0$}} 
\setcounter{equation}{0}

The purpose of the present work is to calculate the gravitational black
hole entropy, and find a mechanism by which the entropy is associated
to the boundary of the black hole manifold. In this section, we focus
our attention to the ``rotating" coordinate system $K_0$.

Thermodynamic properties of a black hole can be determined by the
form of the partition function \eq{1.1} \cite{24,25,26,27,28,29}. The
calculation of $Z[\b,\Om]$ is based on the boundary conditions that
define the set of allowed field configurations $\cC_L$, satisfying
the following properties:
\bitem
\item[i)] $\cC_L$ contains black holes with $(m,J$) belonging to a
small region around some $(m,J)_0$,
\vsm\item[ii)] $\b$ and $\Om$ are constant on the boundary, and
\vsm\item[iii)] there exists a boundary term $I_B$, such that
$\tI=I+I_B$ is differentiable on $\cC_L$.
\eitem

In the lowest-order semiclassical approximation around the black hole
configuration, the logarithm of the partition function takes the form
$$
\ln Z[\b,\Om]=-\tI_{\rm bh}\, ,
$$
where $\tI_{\rm bh}$ is the improved Mielke-Baekler action $\tI$,
evaluated at the black hole \eq{2.8}. On the other hand, using the
general form of the partition function, we obtain the relation
\be
\tI_{\rm bh}=\bar\b (E-\mu M)-S\, ,                        \lab{3.1}
\ee
where $\bar\b=1/T$ is the inverse temperature, $\mu$ is the
chemical potential corresponding to the angular momentum $M$, and
$S$ is the black hole entropy. Thus, the essential step in our
calculation of the black hole entropy is to find $\tI_{\rm bh}$.

\subsection{Regularization}

Before we start calculating $\tI$, let us observe that the value of
the action \eq{2.1} at the black hole configuration is {\it
divergent\/} (Appendix A):
\be
I_{\rm bh}\approx
  \frac{4\pi a\b}{\ell^2}\left(r_\infty^2-r_+^2\right)\, , \lab{3.2}
\ee
where $r_\infty\to\infty$, and $\approx$ denotes an on-shell
equality. Note that this result is of the same form as in \grl\
\cite{29}. One can define a natural regularization procedure by
subtracting the value of $I_{\rm bh}$ at the black hole vacuum, where
$m=J=0$ (see also \cite{30}). The regularized action reads
\be
\Ir=I-\frac{4\pi a\beta}{\ell^2}r^2_\infty \, ,            \lab{3.3}
\ee
and its value at the black hole configuration is finite:
$\Ir\approx-{4\pi a\b}r_+^2/\ell^2$.

Our approach to the black hole thermodynamics relies on the
construction of the improved action $\tI=I+I_B$, in accordance with the
adopted boundary conditions i)--iii). This construction is now
modified by a new element---the regularization procedure. In order to
be sure that the regularization does not spoil the essence of our
approach, we have to verify its consistency with the structure of
boundary terms.

\subsection{Boundary terms}

Now, we wish to improve the form of $I$ on the set of allowed field
configurations $\cC_L$, so that the improved action corresponds to
the grand canonical ensemble. The boundary terms in $\tI=I+I_B$ are
constructed to cancel the unwanted surface terms in $\d I$, arising
from integrations by parts. In other words, $I_B$ is defined by the
requirement $\d(I+I_B)\approx 0$.

The general variation of the action \eq{2.1} at fixed $r$ has the
following form:
\be
\d I\Big|^r=-\left[2a\int b^i\wedge\d\om_i
            +\a_3\int \om^i\wedge\d\om_i
            +\a_4\int b^i\wedge\d b_i\right]^r \, .        \lab{3.4}
\ee
The black hole manifold is taken to be a solid torus with {\it two
boundaries\/}: one at infinity, and one at the horizon. After
completing the calculation, we find out, in contrast to the
Riemannian \grl \cite{29}, that the boundary at the horizon $B^{r_+}$
is {\it absolutely necessary\/}, otherwise the result for the black
hole entropy would be incorrect.

\prg{Spatial infinity.} On the boundary $B^\infty$ located at spatial
infinity, the fields $b^i$ and $\om^i$ are restricted to the family of
black hole configurations \eq{2.8} with $w=0$, $\b$ and $\Om$ are
treated as independent parameters, but their ``on-shell'' values
$\b=2\pi\ell^2 r_+/(r_+^2+\r_-^2)$ and $\Om=N_\vphi(r_+)$ are used at
the end of calculation, in order to avoid conical singularities
\cite{6,29}. The variation of the action $I$ at infinity is calculated
in Appendix B. The first term in \eq{B1} is just the term needed in the
regularization procedure, so that the complete result can be rewritten
as
\be
\d\Ir\Big|^{r\to\infty}=-\d(\b m)+E\d\b-M\d(\b\Om)\, .     \lab{3.5}
\ee
Consequently,
\bitem
\item[(a)] the regularization procedure is consistent with the
structure of boundary terms at infinity.
\eitem

\prg{The horizon.} Looking at the black hole solution at $r=r_+$,
we find
\bsubeq\lab{3.6}
\bea
&&b^0=0\,,\qquad b^2=r_+d\vphi\, ,                        \nn\\
&&\tom^0=0\,,\qquad\tom^2=-2\pi dt+r_+N_\vphi(r_+)d\vphi\,.\lab{3.6a}
\eea
where we used $\b (r_+/\ell^2+r_+\Om N_\vphi(r_+))=2\pi$.
These relations imply
\be
b^a{_0}=0\,,\qquad \om^a{_0}=-2\pi\d^a_2\qquad (a=0,2)\, . \lab{3.6b}
\ee
\esubeq
After going back to the Schwarzschild coordinates, one finds that the
above conditions coincide with those given in Eq. (5.5) of Ref.
\cite{22}.

By using the relations \eq{3.6} as the boundary conditions at the
horizon, the variation of the regularized action at the horizon has
the form \eq{B2}:
\be
\d\Ir\Big|^{r_+}
  =-2\p^2\a_3\d\left(pr_+-2\frac{\r_-}{\ell}\right)\, .    \lab{3.7}
\ee
Thus, the total variation of the regularized action is:
\bea
\d\Ir&=&\d\Ir\Big|^{r\to\infty}-\d I_{reg}\Big|^{r_+}      \nn\\
     &=&-\d(\b m)+E\d\b-M\d(\b\Om)
        +2\p^2\a_3\d\left(pr_+-2\frac{\r_-}{\ell}\right)\,.\lab{3.8}
\eea
We see that $\Ir$ is not differentiable, but this can be easily
corrected.

\prg{Grand canonical action.} Consider the improved action
\be
\tI=\Ir+\b m-2\p^2\a_3\left(pr_+-2\frac{\r_-}{\ell}\right)\,,\lab{3.9}
\ee
the variation of which has the form
\be
\d\tI=E\d\b-M\d(\b\Om) \, .                                \lab{3.10}
\ee
Since $\d\tI$ vanishes when $\b$ and $\Om$ are fixed, $\tI$ is
differentiable, and moreover, it represents the grand canonical
action.

\subsection{Entropy}

Once the grand canonical action is constructed, we can easily find
the black hole entropy. The value of the action \eq{3.9} at the black
hole is
\be
\tI_{\rm bh}=-\frac{\pi r_+}{4G}
    -2\pi^2\a_3\left(pr_+-2\frac{\r_-}{\ell}\right)\, .    \lab{3.11}
\ee
The last term in this expression represents the contribution from the
boundary at the horizon. Using the generalized Smarr formula
in Riemann-Cartan spacetime
$$
\b(E-\Om M)\approx\frac{\pi r_+}{4G}
           +2\pi^2\a_3\left(pr_+-2\frac{\r_-}{\ell}\right)\, ,
$$
we easily find that the above result can be rewritten in the form
\be
\tI_{\rm bh}=\b (E-\Om M)-S\, ,                            \lab{3.12}
\ee
where $\b$ and $\Om$ take their ``on-shell'' values, and
\be
S=\frac{2\pi r_+}{4G}
  +4\pi^2\a_3\left(pr_+-2\frac{\r_-}{\ell}\right)\, .      \lab{3.13}
\ee
Comparing this result with the expected form of $\tI_{\rm bh}$, given
by Eq. \eq{3.1}, we come to the following thermodynamic interpretation:
$\b$ is the inverse temperature, $\Om$ is the thermodynamic potential
corresponding to $M$, and $S$ is the black hole entropy. The above
formula for the black hole entropy coincides with the results obtained
in Refs. \cite{22,23}; in particular, it is in perfect agreement with
the first law of black hole thermodynamics.
\bitem
\item[(b)] The boundary at the horizon produces the last term in
\eq{3.11}, and consequently, its contribution to the black hole
entropy is essential.
\eitem
In \grl, where $\a_3=0$, the last term in \eq{3.11} vanishes and the
boundary at the horizon can be safely ignored, as has been observed
in \cite{29}. More generally, this is true whenever the Chern-Simons
term in the action is absent ($\a_3=0$). On the other hand, whenever
$\a_3\ne 0$, even in Riemannian theory ($p=0$) \cite{31}, the
boundary at the horizon yields a nontrivial contribution, and its
presence cannot be disregarded.

Let us stress that these results hold in the ``rotating" coordinate
system $K_0$. In the next section, we will extend our discussion to
$K_w$.

\section{The black hole entropy in {\boldmath $K_w$}} 
\setcounter{equation}{0}

In this section, we show that the black hole entropy remains
unchanged when we generalize our considerations to an arbitrary
coordinate system of the type $K_w$.  In order to clarify the
dynamical role of boundaries, we construct two specific coordinate
systems, in which the complete contribution to the black hole entropy
comes from a single boundary, $B^\infty$ or $B^{r_+}$.

\subsection{Boundary terms and entropy}

\prg{Spatial infinity.} Using the general result \eq{3.4}, the variation
of $\Ir$ at infinity around the black hole solution \eq{2.8} has the
form
$$
\d\Ir\Big|^{r\to\infty}=-\d(\b m)+E\d\b-M\d(\b\Om)
  -\b^2\left[M\left(\frac{1}{\ell^2}-\Om^2\right)+2\Om E\right]\d w\, .
$$
The last term can be simplified by using the ``on-shell'' equality
$$
\b^2\left[M\left(\frac{1}{\ell^2}-\Om^2\right)
          +2\Om E\right]=8\pi^3\a_3\, ,
$$
which leads to
\be
\d\Ir\Big|^{r\to\infty}\approx
   -\d(\b m)+E\d\b-M\d(\b\Om)-8\pi^3\a_3\d w\, .           \lab{4.1}
\ee
Comparing this expression with the result \eq{3.5} valid in $K_0$, we
see that the only difference comes from the last term in \eq{4.1}.

\prg{The horizon} The black hole solution \eq{3.8} at the horizon
$r=r_+$ is given by
\bea
&&b^0=0\, ,\qquad b^2=r_+d\vphi\, ,                        \nn\\
&&\tom^0=0\, ,\qquad
  \tom^2=-2\pi dt+[r_+N_\vphi(r_+)+2\pi w]d\vphi\, .       \nn
\eea
As a consequence, we find that the relations \eq{3.6b} hold for every
$w$. The variation of $\Ir$ at the horizon yields
\be
\d\Ir\Big|^{r_+}
  =-2\p^2\a_3\d\left(pr_+-2\frac{\r_-}{\ell}\right)
   -8\pi^3\a_3 \d w \, .                                   \lab{4.2}
\ee

Since the last, $w$-dependent terms in \eq{4.1} and \eq{4.2} are
equal, their contribution to $\d\Ir$ is canceled. Consequently,
$\d\Ir$ is the same as in \eq{3.8}, the improved (grand canonical)
action is of the form \eq{3.9}, and we end up with the same formula
\eq{3.13} for the black hole entropy, as expected.
\bitem
\item[(c)] The black hole entropy remains the same in every coordinate
system in $K_w$.
\eitem

\subsection{The analysis of two particular cases}

Now, we wish to analyze the isolated contributions coming from the
boundaries $B^\infty$ and $B^{r_+}$, in two particular coordinate
systems.

\prg{1.} If we chose the parameter $w$ so that
\be
\left(pr_+-2\frac{\r_-}{\ell}\right)+4\pi w=0\, ,          \lab{4.3}
\ee
the variation of $\Ir$ at the horizon vanishes, and the complete
variation is determined by the boundary at infinity. This implies that
both the improved action and the entropy are completely determined by
the contributions from $B^\infty$.
\bitem
\item[(d)] In the coordinate system defined by the condition \eq{4.3},
the complete contribution to the black hole entropy is determined by
the boundary at infinity.
\eitem
In this sense, the boundary $B^{r_+}$ is superfluous and can be
ignored.

This result explains the mechanism used in \cite{29} for \grl\ (in
the ``rotating" coordinate system), and extends it to the more
general Mielke-Baekler model. It tells us that the contribution of
the complete boundary can be effectively reduced just to $B^\infty$,
which is an effect inseparably connected with the specific coordinate
system.

The effect just described may help us to better understand the
relation between (i) the present approach based on the gravitational
partition function, and (ii) the approach based on the Cardy formula
\cite{8,23}. Namely, it seems that the latter approach needs only one
boundary, the boundary at infinity, where all elements of the Cardy
formula are calculated. However, such an assumption would lead to
problems with physical interpretation (see, for instance, the last
reference in \cite{1}). It is not clear that the existence of one
boundary in (ii) is a genuine geometric fact. It might be the result
of an effective description, related, for instance, to the specific
choice of coordinates. Without having a deeper geometric and physical
understanding of the Cardy formula, we cannot properly compare the
geometric content of (i) and (ii).

\prg{2.} Alternatively, we can choose $w$ so that the variation of $\Ir$
vanishes at infinity, for fixed $\b$ and $\Om$:
\be
\b m+8\pi^3\a_3 w=0\, .                                    \lab{4.4}
\ee
The complete variation of $\Ir$ is now determined by the boundary at
the horizon, while $B^\infty$ can be effectively ignored.
\bitem
\item[(e)] In the coordinate system defined by the condition \eq{4.4},
the complete contribution to the black hole entropy is determined by
the boundary at the horizon.
\eitem
It should be noted that the condition \eq{4.4} cannot be realized in
\grl, where $\a_3=0$.

There are arguments that the most natural location for the dynamical
degrees of freedom of the black hole is the horizon \cite{32}.
Clearly, one should ensure that any realization of such an idea
is based on genuine geometric considerations.

\section{Concluding remarks} 
\setcounter{equation}{0}

We investigated thermodynamic properties of the black hole with
torsion using the covariant form of the action in the grand canonical
partition function.

(1) The regularization procedure, needed for a consistent treatment
of the divergent value of action at the black hole configuration, is
shown to be consistent with the boundary conditions corresponding to
the grand canonical partition function.

(2) According to the calculations in the standard coordinate system
$K_0$, the expression for the black hole entropy has the correct
value provided the black hole manifold has not only the boundary at
infinity, but also the boundary at the horizon. The value of the
black hole entropy remains the same in every coordinate system $K_w$.

(3) In the specific coordinate system \eq{4.3}, the complete
contribution to the black hole entropy stems from the boundary at
infinity. This mechanism explains the nature of the corresponding
result in \grl\ \cite{29}. Moreover, it suggest that a similar
analysis of the Cardy formula could help us to properly understand
the underlying boundary geometry, and verify its consistency with the
present approach.

(4) Similarly, the complete contribution in the coordinate system
\eq{4.4} stems from the boundary at the horizon. Such a coordinate
system cannot be realized in \grl, where $\a_3=0$.

\section*{Acknowledgments} 

This work was supported in part by the Serbian Science Foundation,
Serbia.

\appendix
\section{The value of the action} 
\setcounter{equation}{0}

Using the field equations \eq{2.2}, the values of the four pieces of
the action \eq{2.1} at the black hole configuration \eq{2.8} are
given as follows:
\bea
&&I_1=q\ve_{ijk}\int b^i\wedge b^j\wedge b^k
     =6q\pi\b r^2\Big|^{r_\infty}_{r_+}\, ,                \nn\\
&&I_2=-\frac{1}{3}\ve_{ijk}\int b^i\wedge b^j\wedge b^k
     =-2\pi\b r^2\Big|^{r_\infty}_{r_+} \, ,               \nn\\
&&I_3=\frac{1}{2}q\ve_{ijk}\int\om^i\wedge b^j\wedge b^k
      -\frac{1}{6}\ve_{ijk}\int\om^i\wedge\om^j\wedge\om^k
     =pq\pi\b r^2\Big|^{r_\infty}_{r_+}\, ,                \nn\\
&&I_4=\frac{1}{2}p\ve_{ijk}\int b^i\wedge b^j\wedge b^k
     =3p\pi\b r^2\Big|^{r_\infty}_{r_+}\, .                \nn
\eea
The value of the complete action reads:
\be
I_{\rm bh}=(6aq-2\L+\a_3 pq +3\a_4 p)\pi\b r^2\Big|^{r_\infty}_{r_+}
 =\frac{4a\pi\b}{\ell^2}\left(r_\infty^2-r_+^2\right)\, ,  \lab{A1}
\ee
where we used the identities \eq{2.3} and the relation
$q-p^2/4=1/\ell^2$. The result is the same for every coordinate system
in $K_w$.

\section{Variation of the action} 
\setcounter{equation}{0}

The variation of the action \eq{2.1} around the black hole
configuration \eq{2.8} produces two boundary terms, one at infinity and
one at the horizon. The calculation is carried out in the ``rotating"
coordinate system $K_0$.

The contribution from the boundary at infinity is determined by the
relations
\bea
-2a\int b^i\wedge\d\om_i\Big|^{r\to\infty}
  &=&\frac{4\pi a}{\ell^2}r_\infty^2\d\b -\b\d m-J\d(\b\Om)
     +ap Y\, ,                                             \nn\\
-\a_3\int \om^i\wedge\d\om_i\Big|^{r\to\infty}
  &=&(E-m)\d\b -(M-J)\d(\b\Om)
     +\a_3q Y\, ,                                          \nn\\
-\a_4\int b^i\wedge\d b_i\Big|^{r\to\infty}
  &=&\a_4 Y\, ,                                            \nn
\eea
where $Y=2\pi r_\infty^2\d(\b\Om)+\d(\b J)/2a$. Using the second
identity in \eq{2.3}, the sum of these contributions yields
\be
\d I\big|^{r\to\infty}=\frac{4\pi a}{\ell^2}r_\infty^2\d\b
                       -\d(\b m)+E\d\b-M\d(\b\Om)\, ,      \lab{B1}
\ee
which is equivalent to \eq{3.5}.

The contribution from the boundary at the horizon has the form
\eq{3.7}:
\be
\d I\big|^{r_+}=-\a_3\int \om^i\wedge\d\om_i\Big|^{r_+}
     =-2\pi^2\a_3\d\left(pr_+-2\frac{\r_-}{\ell}\right)\, .\lab{B2}
\ee

\end{document}